\documentclass[
reprint,
superscriptaddress,
amsmath,amssymb,
aps,
]{revtex4-2}

\usepackage{graphicx}
\usepackage{dcolumn}
\usepackage{bm}
\usepackage[colorlinks=true,linkcolor=blue,citecolor=blue]{hyperref}%
\usepackage[capitalise]{cleveref}
\newcommand{\numu}{$\nu_{\mu}$}

\begin{document}

\preprint{APS/123-QED}

\title{Trigger-Level Event Reconstruction for Neutrino Telescopes Using Sparse Submanifold Convolutional Neural Networks}

\author{Felix J. Yu}
%  \altaffiliation[Also at ]{Physics Department, XYZ University.}%Lines break automatically or can be forced with \\
\email{felixyu@g.harvard.edu}
\affiliation{Department of Physics and Laboratory for Particle Physics and Cosmology, Harvard University, Cambridge, MA 02138, US}

\author{Jeffrey Lazar}
\email{jlazar@icecube.wisc.edu}
\affiliation{Department of Physics and Laboratory for Particle Physics and Cosmology, Harvard University, Cambridge, MA 02138, US}
\affiliation{Department of Physics and Wisconsin IceCube Particle Astrophysics Center, \\
University of Wisconsin–Madison, Madison, WI 53706, USA}

\author{Carlos A. Arg\"{u}elles}
\email{carguelles@g.harvard.edu}
%  \homepage{http://www.Second.institution.edu/~Charlie.Author}
\affiliation{Department of Physics and Laboratory for Particle Physics and Cosmology, Harvard University, Cambridge, MA 02138, US}

\date{\today}% It is always \today, today,

\begin{abstract}
Convolutional neural networks (CNNs) have seen extensive applications in scientific data analysis, including in neutrino telescopes. However, the data from these experiments present numerous challenges to CNNs, such as non-regular geometry, sparsity, and high dimensionality. Consequently, CNNs are highly inefficient on neutrino telescope data, and require significant pre-processing that results in information loss. We propose sparse submanifold convolutions (SSCNNs) as a solution to these issues and show that the SSCNN event reconstruction performance is comparable to or better than traditional and machine learning algorithms. Additionally, our SSCNN runs approximately 16 times faster than a traditional CNN on a GPU. As a result of this speedup, it is expected to be capable of handling the trigger-level event rate of IceCube-scale neutrino telescopes.
These networks could be used to improve the first estimation of the neutrino energy and direction to seed more advanced reconstructions, or to provide this information to an alert-sending system to quickly follow-up interesting events.

% \begin{description}
% Secondary publications and information retrieval purposes.
% You may use the \texttt{description} environment to structure your abstract;
% use the optional argument of the \verb+\item+ command to give the category of each item. 
% \end{description}
\end{abstract}

\maketitle
% \input{sections/prelude}

%\tableofcontents

\section{Introduction}
\label{sec:intro}

Gigaton-scale neutrino telescopes have opened a new window to the Universe, allowing us to study the highest energy neutrinos.
While there are a variety of proposed designs, many follow the detection principle outlined by the DUMAND project~\cite{RevModPhys.64.259} and consist of an array of optical modules (OMs) deployed in liquid or solid water.
This detector paradigm shows great promise, and analyses by these experiments have already provided the first evidence of astrophysical neutrino sources~\cite{IceCube:2018cha,IceCube:2022der}.
Before they can be analyzed, however, high-energy neutrinos must be isolated from the immense cosmic-ray-muon-induced background.
While a high-energy neutrino may trigger a detector once every few minutes, cosmic-ray muons typically induce a trigger rate on the order of kHz.

\begin{figure}[bht!]
\centering
\includegraphics[width=0.47\textwidth]{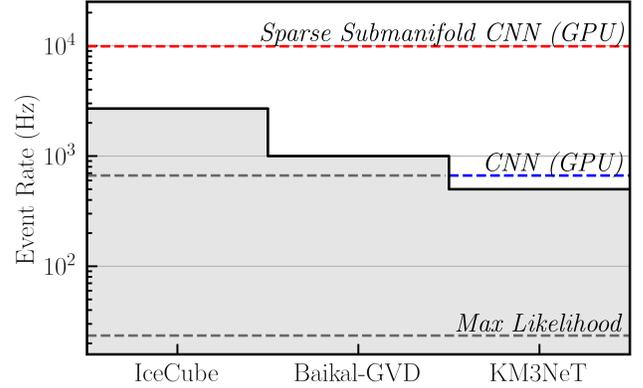}
\caption{\textbf{\textit{Event rates of triggers in different neutrino telescopes~\cite{IceCube_triggers, baikaltdr, km3net_trigger} compared to the run-times of various reconstruction methods.}} Sparse submanifold CNNs and their performance are detailed in this article (the rate shown comes from the angular reconstruction model). The CNN and maximum likelihood method run-times are taken from~\cite{Mirco:2017}. Notably, sparse submanifold CNNs can process events well above standard trigger rates in both ice- and water-based experiments.}
\label{fig:event_rates}
\end{figure}

Since they are unable to traverse a substantial portion of the Earth without coming to rest, cosmic-ray muons have a distinct zenith dependence.
This allows them to be removed by cutting on the reconstructed direction of an event.
Thus, a reliable reconstruction that is capable of keeping up with the $\sim$kHz background rate is the first step in isolating neutrinos.
Moreover, a rapid reconstruction method could serve as part of an alert system that notifies researchers of events that are highly likely to be astrophysical neutrinos.
For example, the real-time follow up of such an IceCube event led to the observation of the first astrophysical neutrino source candidate, TXS 0506+056, by detecting a neutrino in coincidence with a gamma-ray flare~\cite{IceCube:2018cha}.
Along this line, similar efforts are underway in water-based detectors such as ANTARES, see~\cite{Albert:2022oul} for a recent review.

At the trigger-level, a simple but fast reconstruction is typically done by solving a least squares problem via matrix inversion, as is the case for \texttt{LineFit}~\cite{linefit} in IceCube or \texttt{QFit} in ANTARES~\cite{ANTARES:2011vtx}.
Machine learning has shown promise by delivering a comparable quality reconstruction with less runtime requirements~\cite{Mirco:2017,Garcia-Mendez:2021vts}; however, the fastest convolutional neural network (CNN) developed for high-energy neutrinos is not able to keep pace with a kHz-scale trigger-level rate.
In this article, we introduce a reconstruction method using a sparse submanifold CNN (SSCNN), which overcomes this runtime issue.
% We will illustrate our method by focusing on solid-water detectors, but our results and conclusions readily generalize to water-based detectors.
Our SSCNN achieves better angular resolutions than methods such as \texttt{Linefit} while requiring a comparable run-time, enabling improved trigger-level cuts and serving as a better seed for the likelihood-based reconstruction. 
Fig.~\ref{fig:event_rates} summarizes typical event rates found in neutrino telescopes and compares these to the execution rate of various reconstructions. SSCNN is also capable of reconstructing the neutrino energy, a task which has not been done at trigger-level. 

While the rest of this article will concentrate on the implementation of SSCNN in an ice-embedded IceCube-like detector, it should be noted that our method is also applicable to water-based neutrino telescopes, where we expect similar performance gains.
The rest of this article is organized as follows.
In Sec.~\ref{sec:architecture} we motivate and introduce sparse submanifold convolutions; in Sec.~\ref{sec:events} we describe the data sets used for training and testing; in Sec.~\ref{sec:performance} we evaluate the performance of the network.
Finally, in Sec.~\ref{sec:conclusion} we conclude with some parting words.
The code detailing our implementation of SSCNN has been made available at Ref.~\cite{GithubCode}.
\section{Methods}
\label{sec:architecture}

\subsection{Sparse Submanifold Convolutions}

Convolutional neural networks (CNNs) have become the staple architecture for image-like data, and have achieved great success in a wide range of applications, including neutrino physics~\cite{Radovic:2018dip, Aurisano:2016jvx, microboone_cnn, icecube_dnn_reco}. However, data from neutrino telescopes presents inherent challenges to CNNs.
In particular: 
\begin{itemize}
    \item \textit{Non-regular geometry}: CNNs are designed to operate on images, which are arranged on Cartesian grids.  Neutrino telescope sensors are typically spaced irregularly~\cite{KM3NeT:2009xxi, ictdr, baikaltdr, P-ONE:2020ljt}, with varying distances and arrangements in between each sensor.
    % For example, the DOMs in IceCube are roughly arranged in a hexagonal grid, with varying horizontal distances between each of the strings.
    % Additionally, the DeepCore array within IceCube has a different arrangement of DOMs than the rest of the detector.
    \item \textit{Sparsity}: Traditional CNNs use convolutions which operate on all points in the given input data.
    This leads to computational inefficiencies when the data is sparse. 
    \item \textit{High dimensionality}: Events occur in large spatial and temporal scales.
    This makes using traditional CNNs computationally unfeasible on raw 4D data (three spatial and one time) without information loss or significant pre-processing.
\end{itemize}

In this article, we propose a solution to these challenges using sparse submanifold convolutions~\cite{sscnn}.
This strategy has already shown success in liquid argon time projection chamber neutrino experiments~\cite{Abratenko_2021, domine}.
The usage of sparse submanifold convolutions in our network naturally solves the challenges laid out above.
Sparsity and high dimensionality are no longer a concern, as the number of computations performed will depend only on the number of OM hits.
With this improved computational efficiency, we can also handle non-regular geometries more smoothly by using the spatial coordinates of each OM hit (in meters from the center of the detector).
This allows us to consider data of any shape or arrangement, without restricting ourselves to a Cartesian grid; thus our algorithm can be easily adapted from our IceCube-like test case to, \textit{e.g.} IceCube, KM3NeT, P-ONE, \textit{etc}.

\begin{figure}[t]
\centering
\includegraphics[width=0.45\textwidth]{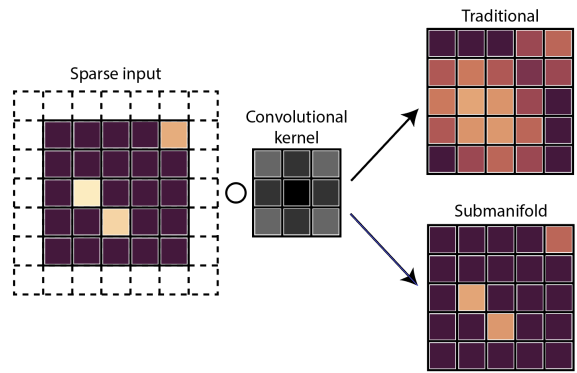}
\caption{\textbf{\textit{Comparison of conventional and submanifold convolution with a Gaussian kernel.}}
The submanifold convolution maintains the sparsity of the input, while the traditional convolution blurs the input, making it less sparse.
In this example, a traditional convolution would require 18 or 25 matrix multiplications for sparse and non-sparse convolution respectively, whereas the bottom image only requires three matrix multiplications.
% It should also be noted the advantage that the sparse conventional gives will diminish with each layer as the number of non-zero-valued pixels``bleed into" zero-valued pixels.
% This same loss will not be seen in sparse submanifold convolution as the number of non-zero-valued pixels is constant from layer-to-layer.
}
\label{fig:compression}
\end{figure}

Our SSCNN replaces traditional convolutions with sparse submanifold convolutions.
While a traditional convolution extracts features by mapping a learned kernel over all input data, a sparse submanifold convolution operates only on the non-zero elements.
This circumvents the inefficiency of using CNNs on sparse data, wherein the vast majority of operations are wasted multiplying zeros together.
Furthermore, to preserve the sparsity of the data after applying multiple layers in succession, sparse submanifold convolutions enforces that the coordinates and number of output activations matches those of the input.
In other words, the features do not spread layer after layer, as shown in Fig.~\ref{fig:compression}.
This is crucial for the efficiency of SSCNNs that are very deep, as the data would otherwise become progressively less sparse throughout the network.
The lack of feature spreading will have a minimal impact on performance as long as the network can rely on local information.It should be noted that SSCNNs still compute over a grid-like structure, but this structure can be arbitrarily large because the network only operates on a submanifold of it.

\subsection{Input Format}
As input, the SSCNN takes in two tensors: a coordinate tensor $C$, and a feature tensor $F$. In symbols:
\begin{equation}
    C = \begin{bmatrix}
    x_1 & y_1 & z_1 & t_1 \\
    \vdots & \vdots & \vdots & \vdots \\
    x_n & y_n & z_n & t_n
    \end{bmatrix}, 
    F = \begin{bmatrix}
    h_1 \\
    \vdots \\
    h_n
    \end{bmatrix},
\end{equation}
where the coordinate tensor is a $n \times 4$ tensor representing the space-time coordinates of the OMs in which there were a nonzero number of photon hits.
The symbol $n$ represents the total number of photon hits in the event, which is variable and can typically range from one to hundreds of thousands.
The feature tensor contains the number of photon hits which occurred within a 1~ns time window on that OM, starting from the time indicated in the coordinate tensor.
Nanosecond units were chosen for very fine timing resolution, as we aim to deploy the network on both low- and high-energy events.
However, depending on the application, the 1~ns time window can be expanded to trade off timing resolution for even better run-time and memory efficiency.

It is worth noting that each coordinate can be associated with any feature vector, offering flexibility for various configurations to encode neutrino telescope data. For instance, we can approach this as a 3-dimensional problem, where we only consider the spatial positions as coordinates and use the timing information as features. However, we decided to treat the timing as coordinates to take advantage of the time structure inherent in neutrino telescope data.
\section{Event Simulation}
\label{sec:events}

\begin{figure*}[t!]
\centering
\includegraphics[width=0.99\linewidth]{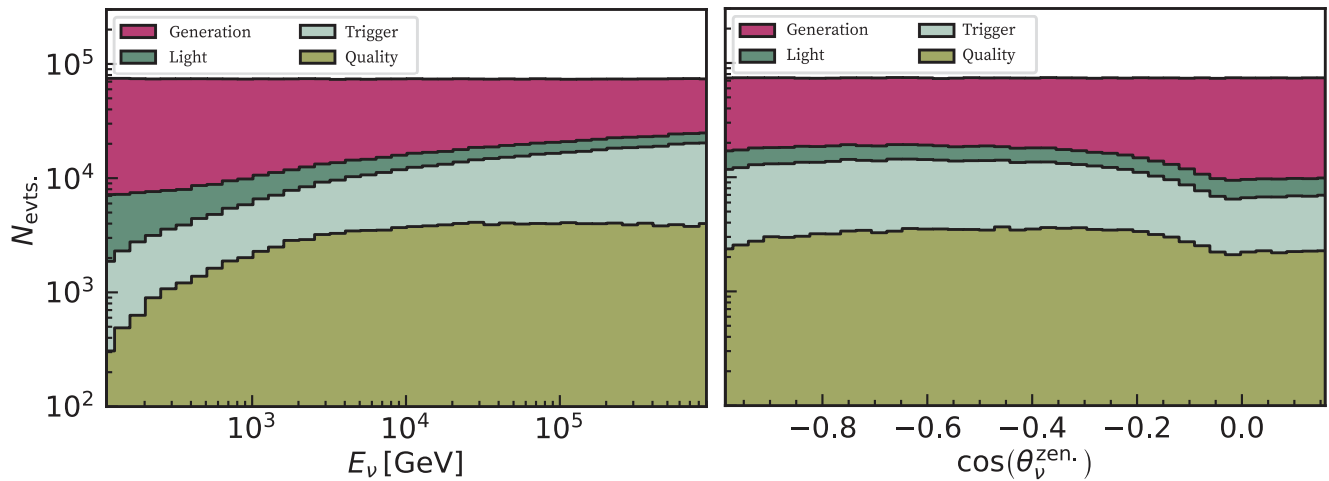}
\caption{\textbf{\textit{Distributions of events in true energy and zenith.}}
\textit{Left:} The distribution of events used as a function of true neutrino energy.
As expected, the generated distribution is flat when binned logarithmically since the generation was sampled according to a $E_{\nu}^{-1}$ distribution.
Furthermore, the fraction of generated events which produce light, and the fraction of light-producing events which pass the trigger threshold increase with energy.
\textit{Right:} The same distribution as a function of true neutrino zenith angle.
Once again the generated distribution is flat in the cosine of this angle, which is proportional to the differential solid angle.
nearly flat, with slightly lower efficiency near the horizon.
}
\label{fig:event_dists}
\end{figure*}

Our benchmark case follows an ice-embedded IceCube-like geometry, where the OMs are spaced our approximately 125 meters horizontally and 17 meters vertically.
The events used in this work are $\mu^{-}$ from \numu{} charged-current interactions.
The initial neutrino sampling, charged lepton propagation, and photon propagation were simulated using the \texttt{Prometheus} package~\cite{santiago_giner_olavarrieta_2022_6804954}.
The incident neutrinos have energies between $10^2$~GeV and $10^6$~GeV sampled from a power-law with a spectral index of -1.
Since most of the events that trigger neutrino telescopes are downward-going cosmic-ray muons, we generated a down-going dataset.
Specifically, the initial momenta have zenith angles between $80^{\circ}$ and $180^{\circ}$.
It is worth noting that this definition of zenith angle is different from the convention which is typically used by neutrino telescopes, which take $0^{\circ}$ to be downgoing.
Internally, \texttt{LeptonInjector}~\cite{IceCube:2020tcq} samples the energy, direction, and interaction vertex.
\texttt{PROPOSAL}~\cite{koehne2013proposal} then propagates the outgoing muon, recording all energy losses that happen within 1~km of the instrumented volume.
\texttt{PPC}~\cite{PPCStandAlone} then generates the photon yield from the hadronic shower and each muon energy loss, and propagates these photons until they either reach an OM or are absorbed.
If a photon reaches an OM, the module ID, module position and time of arrival are recorded.
%Since Antarctic glacier ice properties are not readily available in digital format, we use an approximate ice model that considers the absorption and scattering length given in Ref.~\cite{}

We then add noise in the style of~\cite{Larson:2013xbf} to the resulting photon distributions.
This model accounts for nuclear decays in the OMs' glass pressure housings and thermal emission of an electron from a PMT's photocathode.
The latter process strongly depends on the ambient temperature near the OM and varies between PMTs.
Since this information is not publically available, we simplify the model and vary the thermal noise rate linearly from 40~Hz at the top of the detector to 20~Hz near the bottom, which approximately agrees with the findings from~\cite{Larson:2013xbf}.
We then take the nuclear decay rate to be 250~Hz and generate a number of photons drawn from a Poisson distribution with a mean of 8 for each decay.

Before moving on, it is important to note that the photons generated in the previous steps are only tracked to the surface of the OM.
In a full simulation of the detection process, one would need to simulate the electronics inside the OM, which could introduce timing uncertainties.
Furthermore, the digitized signal reported by the \textit{e.g.} the IceCube OMs must be unfolded to get the number of photons per unit time.
These steps require access to proprietary information that is not available externally.
Thus, we cannot include the effects from these detail detector performance in our simulation.

For example, the process by which IceCube unfolds the photon arrival times is described in~\cite{IceCube:2013dkx}.
They find that this process introduces a timing uncertainty typically on the order of 1~ns but that may grow up 10~ns under certain conditions.
While this may affect our results, we expect the impact to be small since by group the photon arrival times into ns-wide bins, we are introducing a timing uncertainty with a similar scale.

Once all photons have been added, we then implement a trigger criteria similar to the one described in~\cite{IceCube:2016zyt}.
This requires that a pair of neighboring or next-to-neighboring OMs see light in a 1~$\mu$s time window.
If 8 such pairs are achieved in a 5~$\mu$s time window, we consider the trigger to be satisfied.
As before, the exact details of the triggering process require access to proprietary information; however, the events which pass our trigger should be qualitatively similar to those which would trigger IceCube.
After this cut, we are left with 462892 events from 3 million simulated events, which we split between the training and test data sets of 412892 and 50000 sizes respectively.
One can see distributions of the events which pass this trigger as a function of true energy, zenith, and azimuth in~\cref{fig:event_dists}.

In addition to the trigger-level dataset, we also evaluate the network performance on a dataset with further quality cuts, so that we can understand performance on events which are more likely to make it into a final analysis.
In order to do this, we consider three quantities: $N_{\rm{OM}}$, $r_{\rm{COG}}$, and $R_{\rm{ell}}$.
The first two quantities---the number of distinct OMs that saw light and the distance between the charge-weighted center of gravity and the center of the detector---are fairly straightforward, but the last requires more explanation.
To compute $R_{\rm{ell}}$, we fit a two-parameter ellipsoid to all OMs which saw light, and then take the ratio of the long axis to the short axis.
A ratio close to one indicates a spherical events, whereas larger ratios indicate longer, track-like events.

% \begin{figure*}[bt]
% \centering
% \includegraphics[width=0.99\linewidth]{figures/zenith_energy_combined.pdf}
% \caption{\textbf{\textit{Distributions of events in true energy and zenith.}}
% \textit{Left:} The distribution of events used as a function of true neutrino energy.
% As expected, the generated distribution is flat when binned logarithmically since the generation was sampled according to a $E_{\nu}^{-1}$ distribution.
% Furthermore, the fraction of generated events which produce light, and the fraction of light-producing events which pass the trigger threshold increase with energy.
% \textit{Right:} The same distribution as a function of true neutrino zenith angle.
% Once again the generated distribution is flat in the cosine of this angle, which is proportional to the differential solid angle.
% nearly flat, with slightly lower efficiency near the horizon.
% }
% \label{fig:event_dists}
% \end{figure*}

We perform straight cuts on these variables, requiring events have $N_{\rm{OM}} > 11$, $r_{\rm{COG}} < 400$~m, and $2 < R_{\rm ell} < 8$.
The first cut removes low-charge events which are difficult to reconstruct, while the second removes ``corner clipper'' events caused by $\mu^{-}$ passing near the edge of the detector.
The final cut on $R_{\rm{ell}}$ helps ensure that the events have a long lever-arm for angular reconstruction.

These cuts reduce the split training and testing dataset sizes to 108585 and 13183 events, respectively.
The spatial sparsity of these improved quality events is about $\sim3\%$, as there are 154~OMs that were hit on average, out of the 5,160 total OMs in our example detector.
For the trigger level events, the spatial sparsity is about $\sim2\%$.
The time dimension adds another level of sparsity, as typical events can last tens of thousands of nanoseconds compared to the microsecond time window.
\section{Performance}
\label{sec:performance}

\subsection{Training and Architecture Details}

\begin{figure*}[t]
\centering
\includegraphics[trim=0 500 0 450, clip, width=0.99\textwidth]{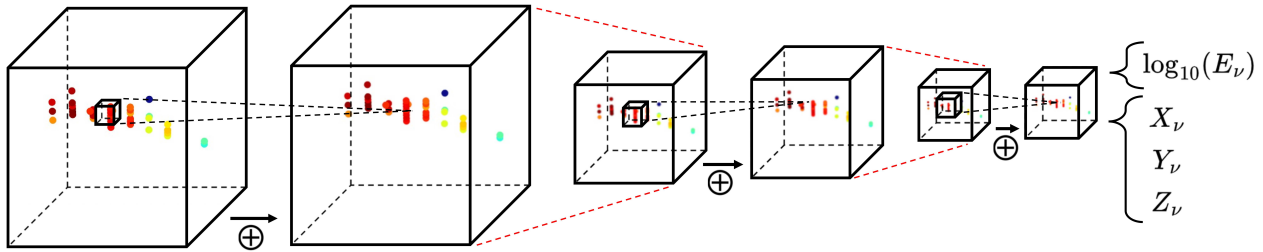}
\caption{\textbf{\textit{Network architecture overview.}} The network accepts as input a 4D point cloud of photon OM hits, as shown by the colored points in the figure. The color indicates the timing (red is earlier, blue is later). Residual connections, denoted with $\oplus$, are used in between convolutions. Downsampling (dashed red lines) is performed after a series of convolutions. The final layer of the network is a fully connected layer, which outputs the predicted quantity. In this work, we train the network to predict the logarithmic $\nu_\mu$ energy, and the three components of the normalized $\nu_\mu$ direction vector.}
\label{fig:arch}
\end{figure*}

We utilize a ResNet-based architecture, taking advantage of residual connections between layers to promote robust learning for deeper networks.
More details on the network architecture can be found in~\cref{fig:arch}.
A typical block of the network consists of a sparse submanifold convolution, followed by batch normalization and the parametric rectified linear unit (PReLU) activation function.
The selection of the activation function was determined after examining prevalent alternatives, such as the conventional ReLU or a smooth approximation, like the SELU.
Downsampling is performed using a stride 2 sparse submanifold convolution.
We use the \texttt{PyTorch} deep learning framework and the \texttt{MinkowskiEngine}~\cite{minkowski} library to implement the network.

This article focuses on the training of three distinct models. The objective of two of these models is the prediction of the three components of the neutrino directional pointing vector ($X_\nu$, $Y_\nu$, $Z_\nu$), with one model trained on the trigger-level dataset and the other trained on the quality dataset. The directional vector is learned rather than the zenith and azimuth angles because of complications with azimuthal periodicity and undesirable boundary condition behavior at large or small angles. Another model was trained to infer the primary neutrino energy $E_\nu$. The network is trained to predict the logarithmic energy, $\log_{10}(E_\nu)$, as they vary over a wide range of magnitudes. Each model was trained for 25 epochs using a batch size of 128 and the \texttt{Adam} optimizer.
The initial learning rate was set at $0.001$ and was reduced periodically during training. Dropout and weight decay are employed to prevent overfitting.

% For the purpose of this article, we train the network to infer the primary neutrino energy $E_\nu$, and the three components of its directional pointing vector, ($X_\nu$, $Y_\nu$, $Z_\nu$).
% The directional vector is learned rather than the zenith and azimuth angles because of complications with azimuthal periodicity and undesirable boundary condition behavior at large or small angles.
% The network is trained to predict the logarithmic energy, $\log_{10}(E_\nu)$, and the normalized directional vectors, as they can vary over a wide range of magnitudes. 

To train the energy reconstruction task, the \texttt{LogCosh} loss function is used, since it is more robust to outliers than the standard \texttt{MSE} loss.
The loss function is defined as follows,
\begin{equation}
    \mathcal{L}_E = \frac{1}{N} \sum^N_i{\log{(\cosh{(x_i - y_i)})}},
\end{equation}
where $N$ is the number of events in the batch, $x_i$ are the predictions, and $y_i$ are the labels.
For the angular reconstruction, an angular distance loss function is used, namely,
\begin{equation}
    \mathcal{L}_A = \frac{1}{N} \sum^N_i{\arccos{\left(\frac{\vec{X_i} \cdot \vec{Y_i}}{||X_i|| \: ||Y_i||}\right)}},
\end{equation}
where $\vec{X_i}$ and $\vec{Y_i}$ are the predicted and true directional vectors, respectively.
% Then, the total loss is given by
% %
% \begin{equation}
%     \mathcal{L}_{tot} = (1 - \alpha)\mathcal{L}_E + \alpha\mathcal{L}_A,
% \end{equation}
% where a weighting factor $\alpha$ is applied to each of the separate loss terms. An ensemble of networks was trained from scratch while $\alpha$ was varied. Performance on both energy and angular reconstruction was tested for each of these networks to determine the optimal value of $\alpha$. We found that setting $\alpha = 0.7$ results in superior angular reconstruction performance, without significantly affecting the energy reconstruction. 

% In addition to the SSCNN, we also train a separate CNN on the same datasets that allows for feature spreading. This network would produce the same results as a traditional CNN and serves as a comparison point. However, due to memory constraints, we modified the architecture of this model by slightly reducing its depth and complexity. Additionally, the traditional CNN model can only fit a batch size of 2 during training, which hinders its learning and makes it significantly slower to train. These compromises can allow the SSCNN to outperform the traditional CNN in certain tasks. 

\subsection{Run-time Performance}

\renewcommand{\arraystretch}{1.25}
\begin{table}[bht]
    \centering
    \begin{tabular}{|p{4.25cm}|p{3cm}|}
        \hline 
        \textbf{SSCNN Angular (GPU)} & \textbf{0.101 $\pm$ 0.003 ms} \\
        \hline
        \textbf{SSCNN Energy (GPU)} & \textbf{0.103 $\pm$ 0.008 ms} \\
        \hline
        \textbf{SSCNN Angular (CPU)} & \textbf{37.7 $\pm$ 53.4 ms} \\
        \hline
        \textbf{SSCNN Energy (CPU)} & \textbf{30.6 $\pm$ 48.9 ms} \\
        \hline
        % CNN (GPU) & 14.7 $\pm$ 7.81 ms \\
        % \hline
        Likelihood Angular (CPU) & 36 $\pm$ 152 ms \\
        \hline
        Likelihood Energy (CPU) & 6.58 $\pm$ 23 ms \\
        \hline
    \end{tabular}
    \caption{\textbf{\textit{Per-event average run-time performance.}} The forward pass run-times (mean $\pm$ STD) for SSCNN was evaluated on trigger-level events. A likelihood-based method for energy and angular reconstruction was included for reference~\cite{Mirco:2017}. It should be noted that likelihood-based methods usually require seeding or initial estimates, meaning the actual runtime is longer.}
    \label{tab:runtime}
\end{table}

We evaluate the run-time performance of SSCNN in terms of the forward pass duration on both CPU and GPU hardware.
The CPU benchmark is performed on a single core of an Intel Xeon Platinum 8358 CPU, while the GPU benchmark uses a 40~GB NVIDIA A100. As is generally the case for neural networks, running on GPU is preferred due to its superior parallel computation capabilities. Additionally, the use of sparse submanifold convolutions has greatly enhanced the GPU memory efficiency, enabling us to run larger batch sizes during inference. SSCNN can reconstruct direction at a rate of 9901 Hz on a 40~GB NVIDIA A100 GPU, while handling a batch size of 12288 events simultaneously.
This is fast enough to handle the expected $\sim$kHz current and planned large neutrino telescopes.
% This rate is over four times that of the current IceCube trigger rate.

The run-time on a single-core CPU is slower and largely dependent on the number of photons hits in the event due to the limited parallel computation capabilities. However, SSCNN run-time on a CPU core is comparable to that of the likelihood-based angular method and is more consistent, as indicated by the lower standard deviation on the run-time distribution. The run-time results on both GPU and CPU are summarized in Table~\ref{tab:runtime}.

% One of the most significant benefits of switching to sparse submanifold convolutions is their improved GPU memory efficiency.
% This enables larger batch sizes during inference, taking advantage of parallel computation.
% On a 40~GB NVIDIA A100~GPU, the SSCNN can process a batch size of 12,288 events at a time.
% We also evaluate the network runtime performance on a single core of an Intel Xeon Platinum 8358 CPU.
% The runtime on a single-core CPU is slower and largely dependent on the number of photons hits in the event due to the limited parallel computation capabilities.
% Because of this, running on GPU is preferred, as is generally the case for neural networks.

\subsection{Reconstruction Performance}

\begin{figure}[t!]
    \centering
    \includegraphics[width=0.48\textwidth]{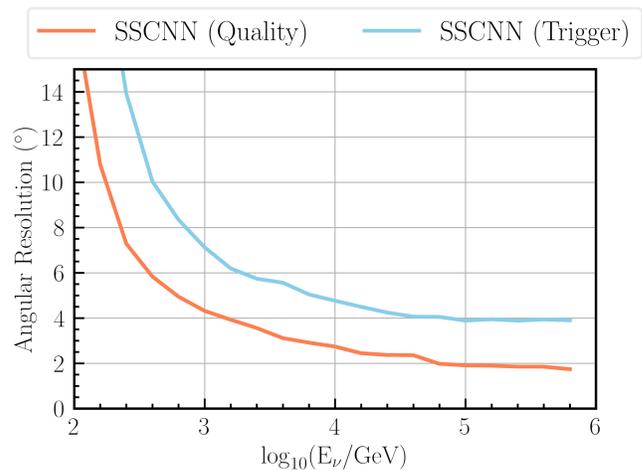}
    \caption{\textbf{\textit{Angular reconstruction performance as a function of the true neutrino energy.}} The angular resolution results are binned by the true neutrino energy, with the median taken from each bin to form the lines shown.
    }
    \label{fig:angular_results}
\end{figure}

\begin{figure}[t!]
    \centering
    \includegraphics[width=0.48\textwidth]{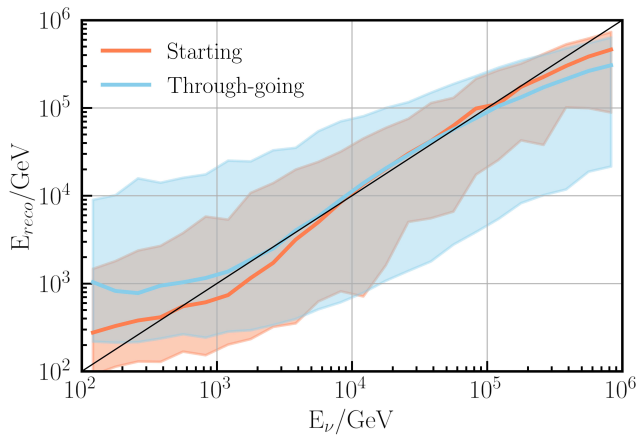}
    \caption{\textbf{\textit{Energy reconstruction performance at the trigger-level.}} The solid lines show the median of the predicted $\log_{10}(E_{\nu})$, while the shaded regions are the $5\%$ to $95\%$ confidence level bands. The events in the test dataset are separated into starting and through-going events. The solid black line serves as a reference for a perfect reconstruction.
    }
    \label{fig:energy_results}
\end{figure}

We first test SSCNN on reconstructing the direction of the primary neutrino.
We measure performance using the angular resolution metric, which is calculated by taking the angular difference between the predicted and true directional vectors.
Fig.~\ref{fig:angular_results} shows the angular resolutions as a function of the true neutrino energy.
Lower-energy events generally produce less photon hits, leading to a shorter lever-arm and, consequently, worse resolution.
As expected, the trigger-level events are harder to reconstruct due to the lower light yield and the presence of corner-clipper events.
On this event selection, SSCNN is able to reach under 4\textdegree{} median angular resolution on the highest-energy events.
% However, poorly-defined events, such as the corner-clippers previously mentioned, can easily pass the SMT-8 threshold.
% These events have little directional information to provide for the network due to their morphology.
Enforcing the previously described quality cuts improves the results of the SSCNN by roughly 2\textdegree{} across the entire energy range.
This performance is comparable to or better than current trigger-level reconstruction methods used in neutrino telescopes.
For example, the current trigger-level direction reconstruction at IceCube is done using the traditional \texttt{Linefit} algorithm~\cite{linefit}, which has a median angular resolution of approximately $10$\textdegree{} on raw data. 

We also test SSCNN on reconstructing the energy of the primary neutrino.
Fig.~\ref{fig:energy_results} summarizes the energy reconstruction results.
Events where the interaction point of the neutrino occurs outside the detector, known as through-going events, make up the majority of our dataset.
As a result, predicting the neutrino energy has an inherent, irreducible uncertainty produced by the unknown interaction vertex and the muon losses outside of the detector.
This missing-information problem leads to an intrinsic uncertainty in the logarithmic neutrino energy of approximately $0.4$ for a through-going event.
The behavior observed between 100 GeV and 1 TeV is due to the muon being in the minimally ionizing regime, up to around 700 GeV. 
Additionally, the network has a tendency to overpredict at the lowest energies, and underpredict at highest energies.
This can be attributed to the artificial energy bounds on the simulated training dataset. 

\section{Conclusions}
\label{sec:conclusion}

In this article, we have demonstrated the application of an SSCNN for event reconstruction on neutrino telescopes.
We have shown that these networks are capable of maintaining competitive performance on the tasks of energy and angular reconstruction while running on the $\mu$s time scale.
The speedup enables the SSCNN to process events at a rate well above that of the current neutrino telescopes trigger rate, which is expected to be representative of other neutrino telescopes currently operating or under construction, such as IceCube, KM3NeT, P-ONE, and Baikal-GVD.
Reaching this threshold makes the SSCNN a feasible option for online reconstruction at the detector site where resources are limited and where first guesses of the energy and direction of the neutrino are made.
As discussed in the introduction, this can have a substantial impact on current real-time analyses, where our first estimations can also be utilized in an alert-sending system, which will notify collaborators if the detector sees an interesting event. 
Additionally, these reconstructions can serve as seeds for more time-consuming reconstructions, and thus improving these first estimations will be beneficial to all subsequent analyses.

As a promising future direction, the exploration of training SSCNN for other event reconstruction challenges, such as morphology and particle classification, holds great potential. Such tasks could significantly benefit from the accelerated runtime provided by SSCNN.
\acknowledgements
We thank Tong Zhu, Alfonso Garcia Soto, and Miaochen Jin for useful discussions.
Additionally, we would like to thank David Kim for help with \texttt{Linefit} and the rest of the \texttt{Prometheus} authors.
CAA, JL, FJY are supported by the Faculty of Arts and Sciences of Harvard University.
Additionally, FJY is supported by the Harvard Physics Department Purcell Fellowship. 

% \input{sections/alt_introduction}
% % \input{sections/introduction}
% \input{sections/architecture}
% \input{sections/events}
% \input{sections/performance}
% \input{sections/conclusion}
% \input{sections/acknowledge}

% Change style so that references come in order of appearance
\bibliographystyle{unsrt}
\bibliography{ic_ssnet}

\begin{thebibliography}{10}

\bibitem{RevModPhys.64.259}
Arthur Roberts.
\newblock The birth of high-energy neutrino astronomy: A personal history of
  the dumand project.
\newblock {\em Rev. Mod. Phys.}, 64:259--312, Jan 1992.

\bibitem{IceCube:2018cha}
M.~G. Aartsen et~al.
\newblock {Neutrino emission from the direction of the blazar TXS 0506+056
  prior to the IceCube-170922A alert}.
\newblock {\em Science}, 361(6398):147--151, 2018.

\bibitem{IceCube:2022der}
R.~Abbasi et~al.
\newblock {Evidence for neutrino emission from the nearby active galaxy NGC
  1068}.
\newblock {\em Science}, 378(6619):538--543, 2022.

\bibitem{IceCube_triggers}
Very high-energy gamma-ray follow-up program using neutrino triggers from
  icecube.
\newblock {\em Journal of Instrumentation}, 11(11):P11009, nov 2016.

\bibitem{baikaltdr}
Baikal-GVD Collaboration.
\newblock {BAIKAL-GVD: Gigaton Volume Detector in Lake Baikal}.
\newblock 2012.

\bibitem{km3net_trigger}
Bardo Bakker.
\newblock {Trigger studies for the Antares and KM3NeT neutrino telescopes},
  July 2011.

\bibitem{Mirco:2017}
Mirco Hünnefeld.
\newblock {Online Reconstruction of Muon-Neutrino Events in IceCube using Deep
  Learning Techniques}, 2017.

\bibitem{Albert:2022oul}
A.~Albert et~al.
\newblock {Review of the online analyses of multi-messenger alerts and
  electromagnetic transient events with the ANTARES neutrino telescope}.
\newblock 11 2022.

\bibitem{linefit}
M.G. Aartsen et~al.
\newblock Improvement in fast particle track reconstruction with robust
  statistics.
\newblock {\em Nuclear Instruments and Methods in Physics Research Section A:
  Accelerators, Spectrometers, Detectors and Associated Equipment},
  736:143--149, feb 2014.

\bibitem{ANTARES:2011vtx}
J.~A. Aguilar et~al.
\newblock {A fast algorithm for muon track reconstruction and its application
  to the ANTARES neutrino telescope}.
\newblock {\em Astropart. Phys.}, 34:652--662, 2011.

\bibitem{Garcia-Mendez:2021vts}
J.~Garc\'\i{}a-M\'endez, N.~Gei\ss{}elbrecht, T.~Eberl, M.~Ardid, and S.~Ardid.
\newblock {Deep learning reconstruction in ANTARES}.
\newblock {\em JINST}, 16(09):C09018, 2021.

\bibitem{GithubCode}
Felix Yu and Jeffrey Lazar.
\newblock \url{https://github.com/felixyu7/nt_ssnet}, 2022.

\bibitem{Radovic:2018dip}
Alexander Radovic, Mike Williams, David Rousseau, Michael Kagan, Daniele
  Bonacorsi, Alexander Himmel, Adam Aurisano, Kazuhiro Terao, and Taritree
  Wongjirad.
\newblock {Machine learning at the energy and intensity frontiers of particle
  physics}.
\newblock {\em Nature}, 560(7716):41--48, 2018.

\bibitem{Aurisano:2016jvx}
A.~Aurisano, A.~Radovic, D.~Rocco, A.~Himmel, M.~D. Messier, E.~Niner,
  G.~Pawloski, F.~Psihas, A.~Sousa, and P.~Vahle.
\newblock {A Convolutional Neural Network Neutrino Event Classifier}.
\newblock {\em JINST}, 11(09):P09001, 2016.

\bibitem{microboone_cnn}
R.~Acciarri et~al.
\newblock {Convolutional Neural Networks Applied to Neutrino Events in a Liquid
  Argon Time Projection Chamber}.
\newblock {\em JINST}, 12(03):P03011, 2017.

\bibitem{icecube_dnn_reco}
R.~Abbasi et~al.
\newblock {A Convolutional Neural Network based Cascade Reconstruction for the
  IceCube Neutrino Observatory}.
\newblock {\em JINST}, 16:P07041, 2021.

\bibitem{KM3NeT:2009xxi}
P.~Bagley et~al.
\newblock {KM3NeT: Technical Design Report for a Deep-Sea Research
  Infrastructure in the Mediterranean Sea Incorporating a Very Large Volume
  Neutrino Telescope}.
\newblock 2009.

\bibitem{ictdr}
J.~Ahrens et~al.
\newblock {IceCube Preliminary Design Document}.
\newblock 2001.

\bibitem{P-ONE:2020ljt}
Matteo Agostini et~al.
\newblock {The Pacific Ocean Neutrino Experiment}.
\newblock {\em Nature Astron.}, 4(10):913--915, 2020.

\bibitem{sscnn}
Benjamin Graham and Laurens van~der Maaten.
\newblock Submanifold sparse convolutional networks, 2017.

\bibitem{Abratenko_2021}
P.~Abratenko et~al.
\newblock Semantic segmentation with a sparse convolutional neural network for
  event reconstruction in {MicroBooNE}.
\newblock {\em Physical Review D}, 103(5), mar 2021.

\bibitem{domine}
Laura Domin\'e and Kazuhiro Terao.
\newblock Scalable deep convolutional neural networks for sparse, locally dense
  liquid argon time projection chamber data.
\newblock {\em Phys. Rev. D}, 102:012005, Jul 2020.

\bibitem{santiago_giner_olavarrieta_2022_6804954}
Santiago~Giner Olavarrieta.
\newblock {Prometheus, An Open-Source Simulation of Neutrino Telescopes}, July
  2022.

\bibitem{IceCube:2020tcq}
R.~Abbasi et~al.
\newblock {LeptonInjector and LeptonWeighter: A neutrino event generator and
  weighter for neutrino observatories}.
\newblock {\em Comput. Phys. Commun.}, 266:108018, 2021.

\bibitem{koehne2013proposal}
Jan-Hendrik Koehne, Katharina Frantzen, Martin Schmitz, Tomasz Fuchs, Wolfgang
  Rhode, Dmitry Chirkin, and J~Becker Tjus.
\newblock Proposal: A tool for propagation of charged leptons.
\newblock {\em Computer Physics Communications}, 184(9):2070--2090, 2013.

\bibitem{PPCStandAlone}
Dmitry Chirkin.
\newblock \texttt{PPC} standalone code.
\newblock \url{https://icecube.wisc.edu/~dima/work/WISC/ppc/}, 2020.

\bibitem{Larson:2013xbf}
Michael~James Larson.
\newblock {\em {Simulation and identification of non-Poissonian noise triggers
  in the IceCube neutrino detector}}.
\newblock PhD thesis, Alabama U., Alabama U., 2013.

\bibitem{IceCube:2013dkx}
M.~G. Aartsen et~al.
\newblock {Energy Reconstruction Methods in the IceCube Neutrino Telescope}.
\newblock {\em JINST}, 9:P03009, 2014.

\bibitem{IceCube:2016zyt}
M.~G. Aartsen et~al.
\newblock {The IceCube Neutrino Observatory: Instrumentation and Online
  Systems}.
\newblock {\em JINST}, 12(03):P03012, 2017.

\bibitem{minkowski}
C.~Choy, J.~Gwak, and S.~Savarese.
\newblock 4d spatio-temporal convnets: Minkowski convolutional neural networks.
\newblock In {\em 2019 IEEE/CVF Conference on Computer Vision and Pattern
  Recognition (CVPR)}. IEEE Computer Society, jun 2019.

\end{thebibliography}

\end{document}